# Ferromagnetism in $Ga_{1-x}Mn_xP$: evidence for inter-Mn exchange mediated by localized holes within a detached impurity band

M. A. Scarpulla[1], B. L. Cardozo[1], W. M. Hlaing Oo[2], M. D. McCluskey[2], and O. D. Dubon[1,*]

[1]*Department of Materials Science & Engineering, University of California at Berkeley and Lawrence Berkeley National Laboratory, Berkeley, California 94720*

[2]*Department of Physics, Washington State University, Pullman, WA 99164-2814*

ABSTRACT

We report an energy gap for hole photoexcitation in ferromagnetic $Ga_{1-x}Mn_xP$ that is tunable by Mn concentration ($x \leq 0.06$) and by compensation with Te donors. For $x\sim0.06$, electrical transport is dominated by excitation across this gap above the Curie temperature ($T_C$) of 65 K and by thermally-activated hopping below $T_C$. Magnetization measurements reveal a moment of 3.5 $\mu_B$ per substitutional Mn while the large anomalous Hall signal unambiguously demonstrates that the ferromagnetism is carrier-mediated. In aggregate these data indicate that ferromagnetic exchange is mediated by holes localized in a Mn-derived band that is detached from the valence band.

Diluted magnetic semiconductors (DMSs) are materials where a few atomic percent of a magnetic element is added to a non-magnetic semiconductor. Because of their unique combinations of magnetic and semiconducting properties and their potential for use as both injection sources and filters for spin-polarized carriers, these materials have been suggested for use in spin-based electronics, or *spintronics*. The discovery of ferromagnetism in the III-Mn-V systems $In_{1-x}Mn_xAs$ and $Ga_{1-x}Mn_xAs$ [1,2] ushered in more than a decade of intense experimental and theoretical research [3-6].

There is general consensus that the inter-Mn exchange in $Ga_{1-x}Mn_xAs$ and other ferromagnetic DMSs such as $Zn_{1-x}Mn_xTe$ [7] and $Ga_{1-x}Mn_xSb$ [8] is mediated by holes in extended or weakly localized states [9]. In $Ga_{1-x}Mn_xAs$, a large central-cell potential arising from the Mn 3d states results in both a greater acceptor binding energy (~110 meV) and a more localized ground-state than for shallow acceptors [10]. The character of the mediating holes— valence band-like, (Mn) impurity band-like, or mixed— is central to establishing accurate models for III-Mn-V ferromagnetism; yet this has not been conclusively established even for $Ga_{1-x}Mn_xAs$ [11,12]. With a larger hole binding energy by Mn acceptors (~400 meV) [13], Mn-doped GaP displays greater hole localization and is expected to have stronger p-d hybridization due to its shorter bond length. $Ga_{1-x}Mn_xP$ is thus an important test bed for understanding the interplay between localization and carrier-mediated exchange [14].

In this Letter, we present experimental evidence demonstrating carrier-mediated ferromagnetism in $Ga_{1-x}Mn_xP$ and the presence of a gap separating a Mn-derived band from the valence band. At low temperatures (i.e. < $T_C$), holes are highly localized; despite their strongly insulating nature, $Ga_{1-x}Mn_xP$ films with nominal x~0.06 exhibit $T_C$s above 60 K. Previous reports [15] have found an unusual ferromagnetism in polycrystalline GaP layers containing Mn; however, the origin of the ferromagnetism was not conclusively established.

Samples for this study were prepared by ion implantation followed by pulsed-laser melting (II-PLM), which we have used previously to synthesize $Ga_{1-x}Mn_xAs$ layers with $T_C$ up to 137 K [16-19]. (Note: to date maximum reported $T_C$ for films grown by molecular beam epitaxy and subjected to extended low-temperature furnace annealing is 173 K [20].) Briefly, an unintentionally-doped (n-type, $10^{16}$-$10^{17}$ /cm$^3$) GaP (001) wafer was implanted with 50 keV Mn$^+$ to a dose of $2 \times 10^{16}$/cm$^2$. Each implanted sample was irradiated in air with a single 0.3-0.4 J/cm$^2$ pulse (FWHM = 19 ns) from a KrF excimer laser ($\lambda$ = 248 nm) homogenized to a spatial uniformity of ±5 % by a crossed-cylindrical lens homogenizer. Extended etching in concentrated HCl removed a poorly-regrown surface layer [17]. Ion channeling techniques including particle induced X-ray emission (PIXE) were used to assess the crystalline quality, total Mn dose, and substitutional Mn fraction.

The solid line in Fig. 1 presents the magnetization of a $Ga_{0.94}Mn_{0.06}P$ sample irradiated at 0.4 J/cm$^2$. A linear dependence of magnetization on temperature up to its $T_C$ of 60 K is observed, which can be understood in terms of a small carrier density and/or carrier localization [18,21]. Samples from the same Mn implantation irradiated between 0.3-0.35 J/cm$^2$ show a maximum $T_C$ of 65 K while $Ga_{0.97}Mn_{0.03}P$ films display $T_C$ up to 23 K [17]. At 5 K and in a field of 5 T, the magnetization of the $Ga_{0.94}Mn_{0.06}P$ film saturates at a value corresponding to 3.5 $\mu_B$ per *substitutional* Mn (and 2.3 $\mu_B$ per *total* Mn), reflecting the measured Mn substitutionality of ~67% (and thus an effective composition for the dilute alloy $x_{eff}$ = 0.04). Because of the rapid regrowth from the liquid phase involved in our process, none of the Mn is present as interstitials as determined by PIXE measurements; the remaining 33% of Mn atoms likely exist as small clusters [22]. We interpret the decreased magnetization and $T_C$ from the sample co-implanted with Mn and Te ions as being due to the reduced hole concentration (p) resulting from partial compensation by Te donors. The changes in $T_C$ accompanying variations of x and p are evidence that the ferromagnetism in $Ga_{1-x}Mn_xP$ is due to a carrier-mediated phase.

The filled circles in Fig. 2 represent the sheet resistivity of a $Ga_{0.94}Mn_{0.06}P$ sample with $T_C$ = 65 K as a function of inverse temperature. The sample is clearly insulating and its resistivity is well described by a model of two thermally-activated processes:

$$\rho(T)^{-1} = (C_1 \exp\{E_1/k_BT\})^{-1} + (C_2 \exp\{E_2/k_BT\})^{-1} \qquad (1)$$

The free parameters are the activation energies $E_{1,2}$ and the pre-exponential constants $C_{1,2}$. Fitting the data to this simple model gives the high- and low-temperature activation energies as ~31 and ~6 meV, respectively. In p-type semiconductors, thermally-activated resistivity in the high-temperature range of the extrinsic regime is typically associated with hole transitions between the valence band and bound-acceptor states. Based on this and the spectroscopic data discussed below, we assign the high-temperature ~31 meV activation energy to excitation across a gap between a Mn impurity band and the valence band. The fact that the change in slope occurs near $T_C$ is consistent with the formation of a continuous hopping transport path at a percolation transition of magnetic polarons [23]. The low-temperature 6 meV process is also consistent with these notions; however, other explanations may also be possible.  The open circles represent the resistivity of the sample in an applied field of 7 T, while the lower-right inset shows that the magnetoresistivity ratio, MR = $\rho(7\ T)/\rho(0)-1$, is negative, reaching its largest magnitude of -48 % at the $T_C$ of 65 K.  Figure 3 presents the Hall resistance of the same sample as a function of applied field.  Taking the slope of the 300 K data indicates a hole concentration in the $10^{17}$–$10^{18}$ /cm$^3$ range, the uncertainty being due to a varying Mn composition as a function of depth and the strong anomalous Hall component which has the same sign as in $Ga_{1-x}Mn_xAs$ films.  The magnitude of the anomalous component increases nearly linearly with decreasing temperature below $T_C$, reflecting the magnetization shown in Fig. 1.  These magnetoresistive and anomalous Hall characteristics reflect the intimate relationship between hole transport and ferromagnetism; similar behavior is observed in other III-V ferromagnetic semiconductors [7,8] as well as in manganites [24].

To investigate the hypothesis of a Mn impurity band separated by an energy gap from the valence band, we used Fourier-transform far-infrared photoconductivity (PC) and infrared (IR) absorption spectroscopies.  In PC spectroscopy, excitation of localized holes into extended states results in a change in conductivity that is readily detectable by lock-in techniques with much greater sensitivity than can be achieved with absorption measurements.  The extrinsic PC spectrum from a p-type semiconductor displays a threshold at a photon energy corresponding to the photoexcitation of holes from the neutral acceptor ground state to the valence band; this photoexcitation edge provides a close determination (within ~10%) of the hole binding energy.

The solid line in Fig. 4a shows the far IR photoconductivity response from the Ga$_{0.94}$Mn$_{0.06}$P sample at 4.2 K while the dashed line gives the spectrum incident on the sample. The gross features of the incident spectrum are reproduced in the sample PC spectrum with the exception of the region below ~26 meV. The incident spectrum has appreciable spectral weight in this region; so the delayed onset of the sample response is direct evidence of an energy gap across which carriers are optically excited. This gap energy is in reasonable agreement with the ~31 meV energy deduced from the resistivity data, and together these suggest that the Mn impurity band is distinct from the valence band, as depicted in Fig. 4d.

To further test our hypothesis of a Mn impurity band separated by a gap from the valence band, we measured the photoconductivity spectra of Te-compensated and Ga$_{0.97}$Mn$_{0.03}$P samples. In the Ga$_{0.97}$Mn$_{0.03}$P sample, the onset of the PC response is shifted to a higher energy of ~47 meV, which is as well in reasonable agreement with the measured activation energy of ~53 meV for the high-temperature region of the sample's resistivity. These observations of increased activation energy are consistent with narrowing of the impurity band with decreasing Mn concentration. Figure 4b demonstrates that in the sample compensated with Te donors the onset of the PC response at 4.2 K occurs at ~70 meV, which is indicative of a shift of the Fermi energy into the Mn band due to a reduction of the hole concentration. The spectra taken at higher temperatures display a gradual increase in spectral weight at lower energies and a return of a photoconductivity edge to ~23 meV. At temperatures above 18.5 K the signal increases in intensity but the onset does not shift further to lower energy. These observations are consistent with the thermal redistribution of holes whereby the highest available impurity band (hole) states are occupied sufficiently by 18.5 K for detection via PC. Additionally, the resistivity of this compensated sample rises much faster with decreasing temperature in the high-temperature region, consistent with a larger thermal activation energy.

Figure 4c presents the infrared absorption spectrum in the 150 to 650 meV energy range from the Te-compensated sample at 10 K. The spectrum is very weak and shows a broad feature peaked between 300-400 meV, which is near the Mn acceptor ionization energy in GaP of ~400 meV; we therefore interpret it as resulting from hole transitions from the Mn impurity band to the valence band. These data have been normalized to the spectrum from a sample which was implanted with $Kr^+$ and laser melted under identical conditions; such normalization is used to remove absorption features not directly related to the presence of Mn such as those due to the excitation of GaP phonon modes or those due to the sample processing. The sharp dips in the spectrum reflect non-idealities in the normalization. Nevertheless, the data clearly indicate a rapid decrease of the impurity band density of states approaching the valence band edge, consistent with the PC results.

Models of carrier statistics involving an impurity band centered near 400 meV (with various densities of states) separated by a ~25 meV gap from the GaP valence band yield 300 K free-hole concentrations in the range of $10^{18}$–$10^{19}$ /cm$^3$ and describe the observed temperature dependence of the resistivity at the higher temperatures (above $T_C$). This modeling and the experimentally observed resistivity indicate that the valence-band hole concentration is 1-2 orders of magnitude lower at the $T_C$ of 65 K; it is therefore unlikely that models requiring large concentrations of valence-band holes that are on the order of the Mn concentration of $10^{20}$-$10^{21}$/cm$^3$ can adequately account for the observed ferromagnetism. Our results indicating a lower $T_C$ for $Ga_{1-x}Mn_xP$ than for $Ga_{1-x}Mn_xAs$ of similar Mn composition do not necessarily suggest that the maximum attainable $T_C$ in this system will also be lower; it is possible that stronger ferromagnetism may be achieved by incorporating higher concentrations of Mn via optimization of II-PLM.

The fundamental issue of whether the inter-Mn exchange (and hence $T_C$) changes monotonically across the Ga-Mn-pnictide series [25-27] from $Ga_{1-x}Mn_xSb$ (maximum reported $T_C$ 25 K) [8] to $Ga_{1-x}Mn_xN$ (maximum reported $T_C$ 940 K) [28] is still unresolved. The shorter bond length in GaP should lead to greater p-d exchange between holes and Mn ions than in GaAs. However, this increased exchange energy could contribute to the already significant localization of hole states leading to less overlap of states on different Mn and overall weaker inter-Mn exchange [14]. The presence of a distinct Mn impurity band and the ability to vary the Fermi energy via compensation in $Ga_{1-x}Mn_xP$ enables testing theories of ferromagnetism and spin-polarized transport including the prediction that a change in sign of the anomalous Hall coefficient as the Fermi level moves within an impurity band [29].

In summary, $Ga_{1-x}Mn_xP$ represents a novel DMS alloy system where strongly localized carriers in a detached impurity band stabilize ferromagnetism. The unique electrical, magnetic, and optical properties displayed by this material make it a model system for investigating the rich interplay between bandstructure, carrier localization, and mechanisms of ferromagnetic exchange.

This work was supported by the Director, Office of Science, Office of Basic Energy Sciences, Division of Materials Sciences and Engineering, of the US Department of Energy under Contract No. DE-AC03-76SF00098. The authors thank Y. Suzuki and E.E.

Haller for use of facilities, R. Chopdekar, T.W. Olson, and L.A. Reichertz for experimental assistance, and W. Walukiewicz and K.M. Yu for many fruitful discussions. M.A.S. acknowledges support from an NSF Graduate Research Fellowship. O.D.D. acknowledges support from the Hellman Family Fund.

# REFERENCES

1. H. Ohno et al., Appl. Phys. Lett. **69** (3) 363 (1996).

2. H. Munekata et al., Phys. Rev. Lett. **63** (17) 1849 (1989).

3. H. Ohno, Science **281** 951 (1998).

4. T. Dietl, Semicond. Sci. Technol. **17** 377 (2002).

5. T. Dietl and H. Ohno, MRS Bul. **28** (10) 714 (2003).

6. F. Matsukura, H. Ohno, and T. Dietl, *Handbook of Magnetic Materials*, Vol. 14, ed. K. H. J. Buschow (Elsevier, Amsterdam, 2002), p. 1.

7. T. Dietl et al., in *Proceedings of the NATO Advanced Research Workshop "Recent Trends in Theory of Physical Phenomena in High Magnetic Fields*", eds. I. Vagner, et al. (Kluwer, Dordrecht, 2003), p. 197.

8. E. Abe et. al., Physica E **7** 981 (2000).

9. T. Dietl et al., Science **287** 1019 (2000).

10. M. Linnarsson et al., Phys. Rev. B **55**, 6938 (1997).

11. V.F. Sapega et al., Phys. Rev. Lett. **94**, 137401 (2005)

12. K.S. Burch et al., Phys. Rev. B **70** 205208 (2004) and references therein.

13. B. Clerjaud, J. Phys. C: Solid State Phys. **18** 3615 (1985) and references therein.

14. A.H. MacDonald, P. Schiffer, and N. Samarth, Nature Mater. **4** 195 (2005).

15. N. Theodoropoulou et al., Phys. Rev. Lett. **89** (10) 107203 (2002).

16. M. A. Scarpulla et al., Appl. Phys. Lett. **82** 1251 (2003).

17. M. A. Scarpulla et al., Physica B **340-342**, 908 (2003).

18. M. A. Scarpulla et al., in *Proceedings of the 27th International Conference on the Physics of Semiconductors*, eds. J. Menendez and C.G. Van de Walle, in press.

19. M. A. Scarpulla et al., unpublished.

20. K.Y. Wang et al., in *Proceedings of the 27th International Conference on the Physics of Semiconductors*, eds. J. Menendez and C.G. Van de Walle, in press.

21. S. Das Sarma, E. H. Hwang, and A. Kaminski, Phys. Rev. B **67**, 155201 (2003).

22. K.M. Yu et al., Appl. Phys. Lett. **86** (4) 042102 (2005).

23. A. Kaminski and S. Das Sarma, Phys. Rev. B. **68** 235210 (2003).

24. M. Viret, L. Ranno, and J.M.D. Coey, Phys. Rev. B **55** (13) 8067 (1997).

25. P. Mahadevan and A. Zunger, Appl. Phys. Lett. **85** (14) 2860 (2004).

26. K. Sato et al., J. Magn. Magn. Mater. **272-276** 1983 (2004).

27. T. Jungwirth et al., Phys. Rev. B **66** (1) 012402 (2002).

28. T. Sasaki et al., J. Appl. Phys. **91** 7911 (2002).

29. A.A. Burkov and L. Balents, Phys. Rev. Lett. **91** (5) 057202 (2003).

FIGURE CAPTIONS

**Figure 1** – Low-field (50 Oe) magnetization as a function of temperature for a $Ga_{0.94}Mn_{0.06}P$ film (solid) and a similar sample compensated with Te (dashed).

**Figure 2** – (main) Sheet resistivity, $R_{sheet}$, as a function of inverse temperature for $Ga_{0.94}Mn_{0.06}P$ in zero-field (filled circles) and in a 7 T magnetic field (open circles). The black line through the zero-field data is the fit to the model described in the text. (inset) Magnetoresistive ratio between 7 T and zero-field data as a function of temperature.

**Figure 3** – Hall resistance as a function of magnetic field for $Ga_{0.94}Mn_{0.06}P$ showing the dominance of the anomalous component at low temperatures.

**Figure 4** – (a) Far IR photoconductive response from $Ga_{0.94}Mn_{0.06}P$ sample (solid) and incident spectrum (dashed). An energy gap for photoexcitation is evidenced by the onset of the sample response at ~26 meV. The decreased response of the $Ga_{0.94}Mn_{0.06}P$ sample at ~45 meV arises from absorption in the GaP matrix due to the excitation of optical phonons. (b) Far-IR photoconductive response from the Te compensated sample showing a larger gap and thermal redistribution of holes as temperature increases. (c) Far-IR absorption spectrum from a Te-compensated sample measured at 10 K. The peak centered near 400 meV is due to the Mn impurity band. (d) Schematic density of states showing the Mn impurity band separated from the valence band by an energy gap.

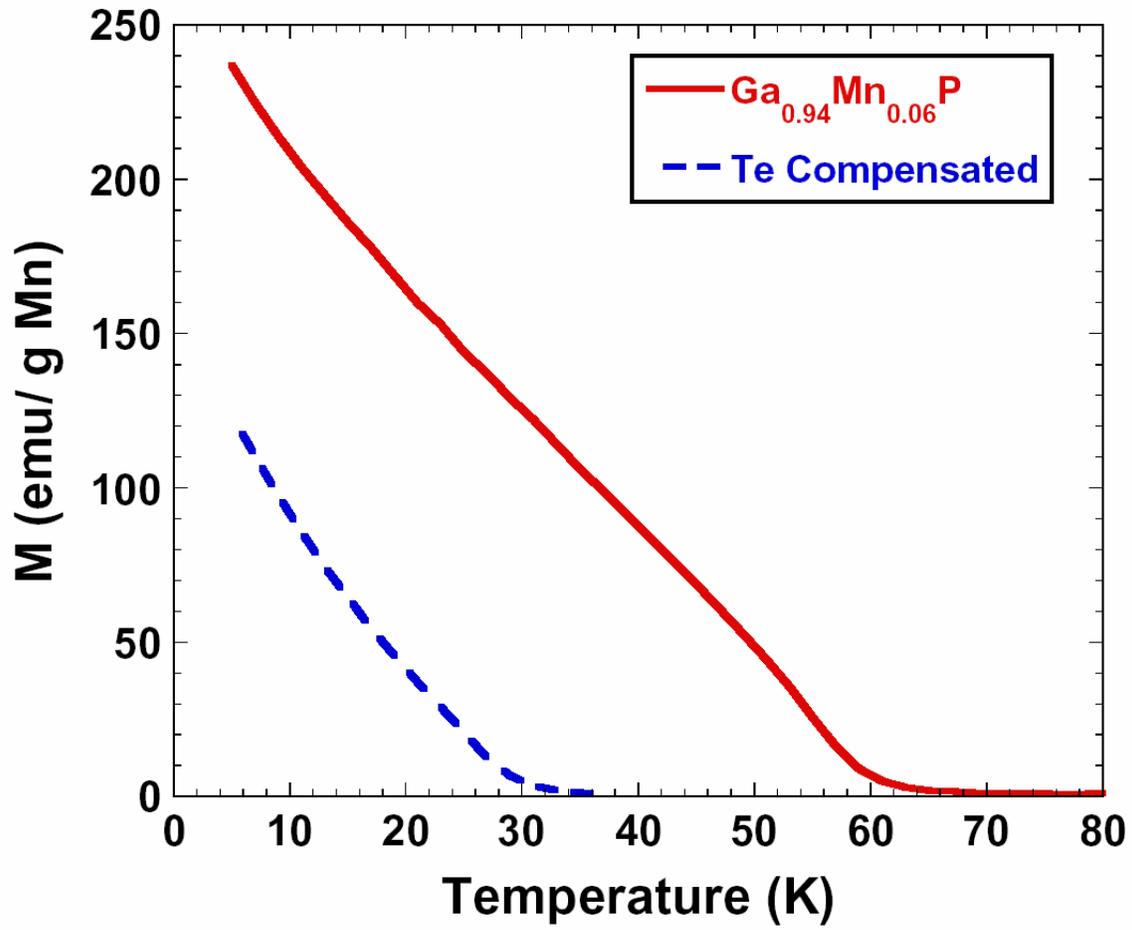

FIGURE 1

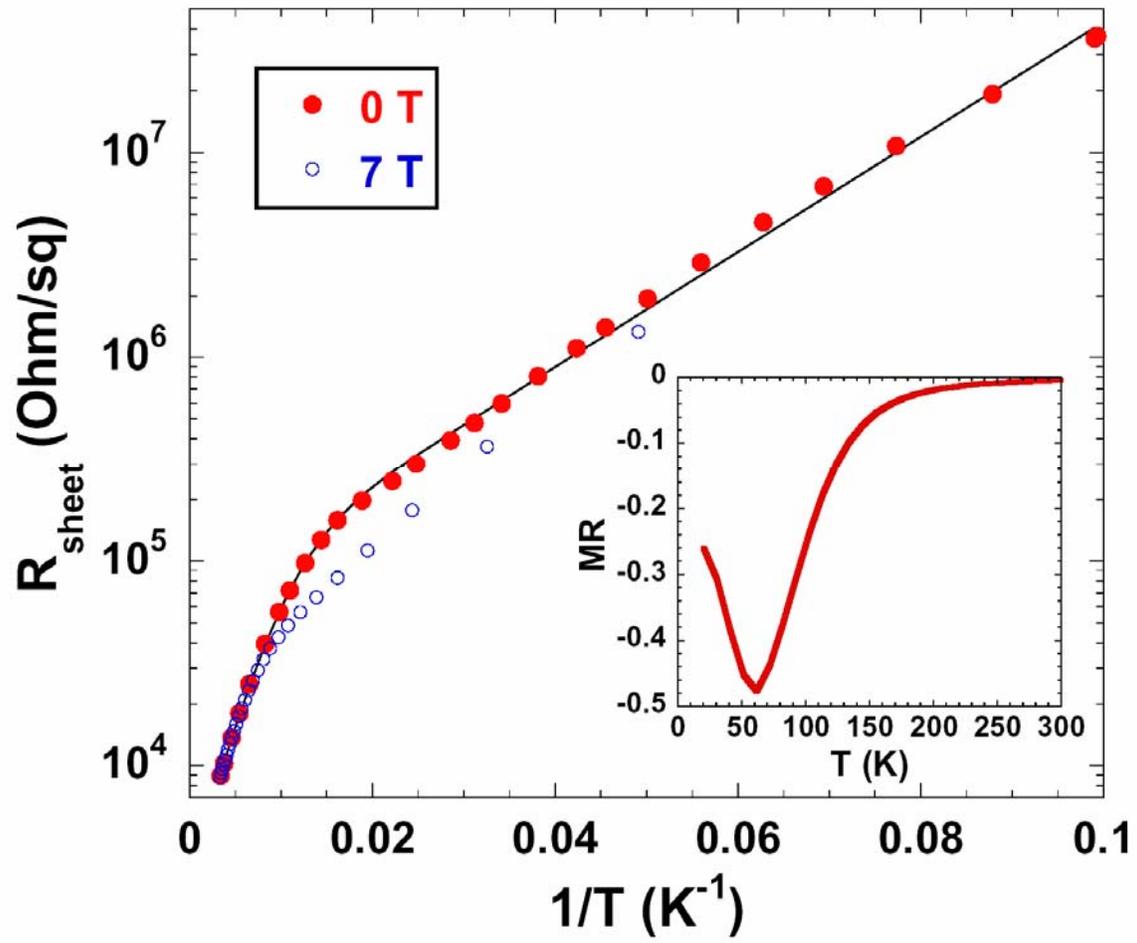

FIGURE 2

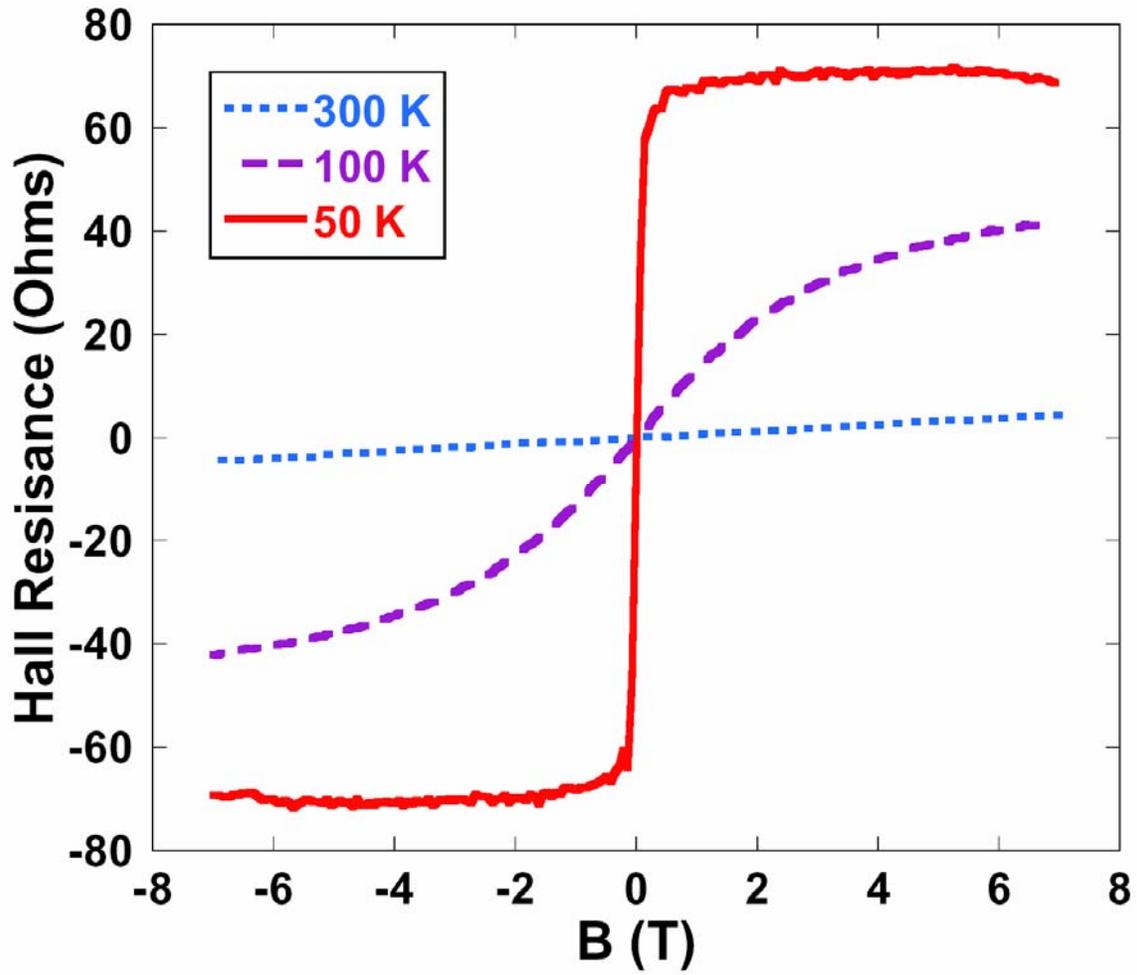

FIGURE 3

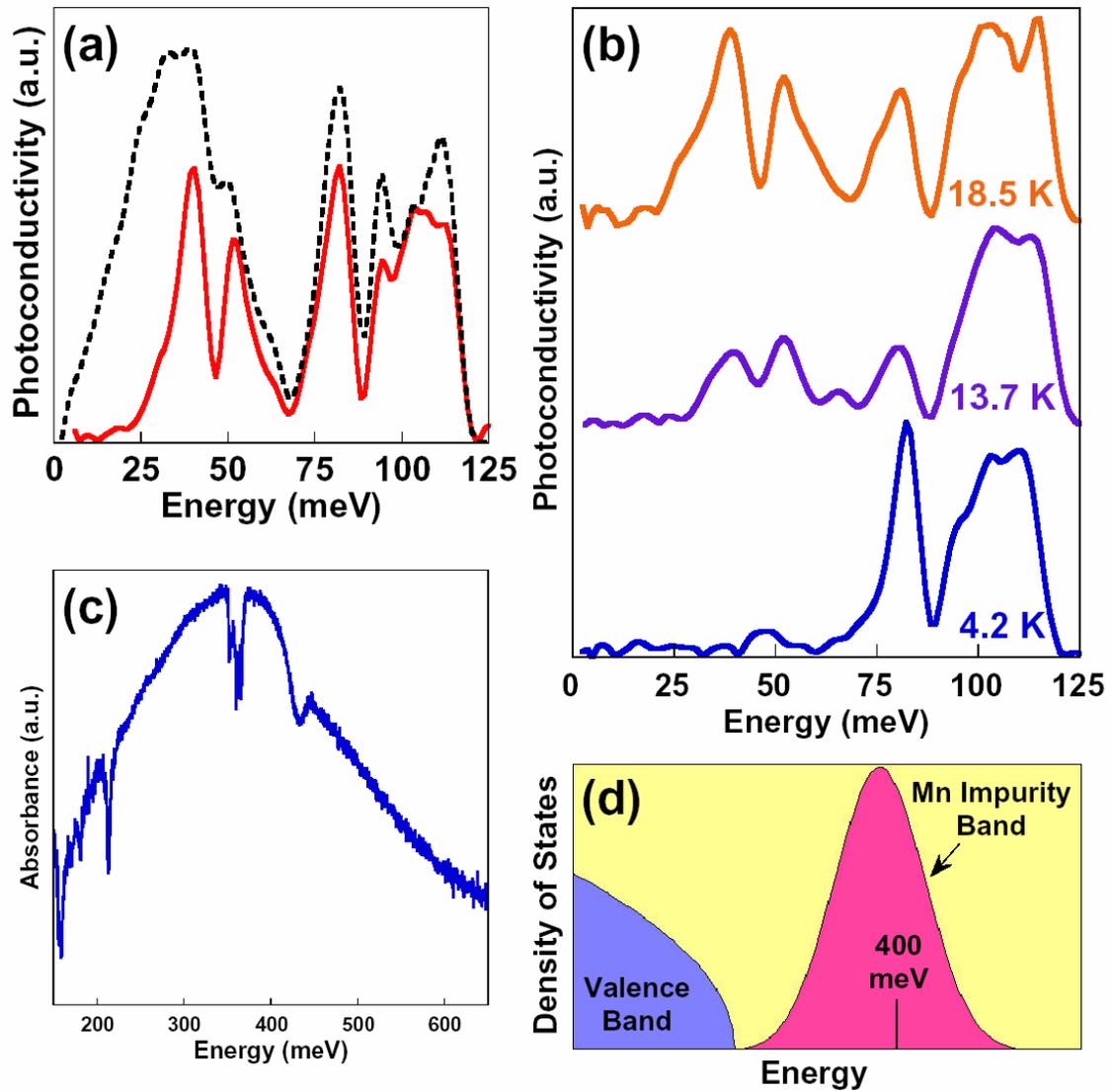

FIGURE 4